\documentclass[useAMS,usenatbib]{mn2e}
\usepackage{epsfig}
\usepackage{amssymb}
\usepackage{amsmath}
\usepackage{lscape}
\usepackage{longtable}
\usepackage[normal]{caption}

\makeatletter

\newcommand{\Rmnum}[1]{\expandafter\@slowromancap\romannumeral #1@}
\makeatother

\title[AGN in dusty hosts: implications for galaxy evolution]{AGN in dusty hosts: implications for galaxy evolution}

\author[M.~Symeonidis et al.]
{\parbox{\textwidth}{\raggedright M.~Symeonidis$^{1}$\thanks{E-mail:
      \texttt{msy@mssl.ucl.ac.uk}}, J.~Kartaltepe$^{2}$,
    M. Salvato$^{3}$, A. Bongiorno$^{4}$, M.~Brusa$^{3}$,
    M. J.~Page$^{1}$, O. Ilbert$^{5}$,
    D. Sanders$^{6}$, A. van der Wel$^{7}$}\vspace{0.4cm}\\
\parbox{\textwidth}{\raggedright $^{1}$ Mullard Space Science Laboratory, University College London, Holmbury St. Mary, Dorking, Surrey RH5 6NT, UK \\
       $^{2}$ National Optical Astronomy Observatory, 950 N. Cherry Ave, Tucson, AZ 85719, US \\
       $^{3}$ Max-Planck-Institut fur extraterrestrische Physik, Giessenbachstrasse 1, 85748 Garching, Germany \\
        $^{4}$ INAF-Osservatorio Astronomico di Roma, via di Franscati 33, 00040 Monte Porzio Catone, Italy \\
       $^{5}$  Laboratoire d’Astrophysique de Marseille, Universite de Provence, CNRS, BP 8, Traverse du Siphon, 13376 Marseille Cedex 12, France \\
$^{6}$  Institute for Astronomy, 2680 Woodlawn Drive, University of Hawaii, Honolulu, HI 96822, USA \\
$^{7}$ Max-Planck-Institute for Astronomy, Koenigstuhl 17, D-69117, Heidelberg, Germany }}

\begin{document}

\date{Accepted  Received; in original form}

\pagerange{\pageref{firstpage}--\pageref{lastpage}} \pubyear{2011}

\maketitle

\label{firstpage}

\begin{abstract}
We present strong empirical evidence for a physical connection between
the occurrence of a starburst (SB) and a luminous AGN phase. Drawing
infrared (IR), X-ray, and optically selected samples from COSMOS, we
find that the locus of type-\Rmnum{2} AGN hosts in the optical colour-magnitude
(U-V/V) and colour-colour (U-V/V-J) space significantly overlaps with
that of IR-luminous ($L_{\rm IR}>$ 10$^{10}$ L$_{\odot}$) galaxies. Based on our observations, we propose that, when simultaneously building their black
hole and stellar masses, type-\Rmnum{2} AGN hosts
are located in the same part of colour-colour space as dusty
star-forming galaxies. In fact, our results show that IR-luminous galaxies at
$z<$1.5 are on average 3 times more likely to host a type-\Rmnum{2} AGN ($L_{\rm X}$\,$>$10$^{42}$\,erg/s) than would be
expected serendipitously, if AGN and star-formation events were
unrelated. In addition, the optical and infrared properties of the AGN/SB
hybrid systems tentatively suggest that the AGN phase might be coeval with a
particularly active phase in a galaxy's star-formation
history. Interestingly, we also find a significant fraction of type-\Rmnum{2} AGN
hosts offset from the dusty galaxy sequence in colour-colour space, possibly representing a
transitional or post-starburst phase in galaxy evolution. Our
findings are consistent with a scenario whereby AGN play a role in the
termination of star-formation in massive galaxies. 
\end{abstract}

\section{Introduction}
\label{intro}

It has long been proposed that black hole growth and starburst activity follow
connected evolutionary paths, inter-regulated through complex feedback
processes and eventually resulting in the observed black hole -- bulge
relationship (e.g. Silk $\&$
Rees 1998\nocite{SR98}; Magorrian et al. 1998\nocite{Magorrian98}). In
the context of extragalactic studies, a popular tool for studying
galaxy evolution is the rest-frame optical colour-magnitude diagram
(CMD) where the position of each galaxy in colour-magnitude space is
linked to its evolutionary state (e.g. Strateva et al. 2001\nocite{Strateva01}; Baldry et al. 2004\nocite{Baldry04};
Weiner et al. 2005\nocite{Weiner05}). The CMD appears bimodal, where
most late-types lie in the blue cloud and most early type galaxies in the red sequence, with the region between the two
(the `green valley'; e.g. Wyder et al. 2007\nocite{Wyder07}) being sparsely populated. The latter observation has led to the conclusion that galaxies spend only a small fraction of
their lifetime in transit from the blue cloud to the red sequence,
which has in turn founded speculations that energetic events
associated with the central accreting black hole (active galactic
nucleus; AGN), could be
responsible for terminating star-formation on short timescales
(e.g. Granato et al. 2004\nocite{Granato04}; Springel, Di Matteo $\&$
Hernquist 2005\nocite{SdMH05}). As a result, the search for evidence
of AGN feedback in extragalactic surveys has been to a large extent
pursued through examination of the properties of green galaxies and
type-\Rmnum{2} AGN hosts. However, these studies have not always been in agreement with
respect to whether green galaxies are a transition
population or whether the colours of type-\Rmnum{2} AGN hosts are an indication of
the quenching of star-formation (e.g. Westoby et al. 2007; Martin
et al. 2007\nocite{Martin07}; Georgakakis et
al. 2008\nocite{Georgakakis08}; Brusa et al. 2009; Brammer et al. 2009\nocite{Brammer09}; Schawinski
et al. 2009\nocite{Schawinski09}, Cardamone et al. 2010\nocite{Cardamone10}). 

Here we aim to address the topic of starburst/AGN synergy, using a sample of
X-ray selected type-\Rmnum{2} AGN and a sample of 70\,$\mu$m-selected galaxies from
the COSMOS 2\,deg$^2$ field. Previous
work on 70\,$\mu$m populations (Symeonidis et al. 2008; 2009;
2010\nocite{Symeonidis08}\nocite{Symeonidis09}\nocite{Symeonidis10};
Kartaltepe et al. 2010a\nocite{Kartaltepe10a}; 2010b\nocite{Kartaltepe10b}), has
shown that the selection at 70\,$\mu$m enables the assembly of a
homogeneous sample of IR-luminous ($L_{\rm IR}$\,$>$10$^{10}$\,L$_{\odot}$) sources, characterised by large
star-formation rates (SFRs). The motivation for this work is to
investigate the link (if any) between black hole accretion and
star-formation, by comparing the location of AGN hosts in the
colour-magnitude, ($U-V$ vs $M_{\rm V}$), and colour-colour ($U-V$ vs $V-J$) diagrams,
(e.g. Rovilos $\&$ Georgantopoulos 2007\nocite{RG07}; Nandra et al. 2007\nocite{Nandra07}; Treister et
al. 2009\nocite{Treister09}; Hickox et al. 2009\nocite{Hickox09};
Silverman et al. 2009\nocite{Silverman09}; Bongiorno et al. 2012\nocite{Bongiorno12}) to that of IR-luminous galaxies (Symeonidis et
al. 2010; Kartaltepe et al. 2010b). 

The paper is laid out as follows: in section \ref{sec:sample} we describe the sample
and in section \ref{sec:cmd} we present our results and
analysis. Our conclusions are reported in section
\ref{sec:conclusions}. Throughout, we adopt a concordance cosmology of
H$_0$=70\,km\,s$^{-1}$Mpc$^{-1}$, $\Omega_{\rm M}$=1-$\Omega_{\rm
  \Lambda}$=0.3.

\section{Observations}
\label{sec:sample}

\subsection{The data}
Our data comes from the 2\,deg$^2$ COSMOS field (Scoville et
al. 2007\nocite{Scoville07}). 
The optical sample consists
of 438,226 objects with $i_{\rm auto}^+$ (Subaru)$<$25 (AB), with 30-band imaging
data combined for the derivation of photometric
redshifts and rest-frame magnitudes (see Capak et
al. 2007\nocite{Capak07}; Ilbert et al. 2009\nocite{Ilbert09}). 
In this work we also take advantage of the 24\,$\mu$m and 70\,$\mu$m \textit{Spitzer}/MIPS observations (Frayer et al. 2009\nocite{Frayer09}; Le Floc'h et al. 2009\nocite{LeFloch09}) acquired as part of the
\textit{Spitzer}  COSMOS legacy survey (Sanders et
al. 2007\nocite{Sanders07}), as well as the $\sim$50\,ks \textit{XMM-Newton} survey of the COSMOS field (Hasinger et al. 2007\nocite{Hasinger07};
Cappelluti et al. 2007\nocite{Cappelluti07};
2009\nocite{Cappelluti09}). The 70\,$\mu$m catalogue used in this work
together with details on the multiwavelength
data cross-matching and in-depth analysis of the 70$\mu$m population
is presented in Kartaltepe et al. (2010a, 2010b). 
The \textit{XMM-Newton} X-ray catalogue is compiled and cross-matched to the optical
sample in Brusa et al. (2010\nocite{Brusa10}). There
are 1550 extragalactic X-ray sources with unambiguous optical counterparts.

  \begin{figure}
   \centering
\begin{tabular}{c}
\epsfig{file=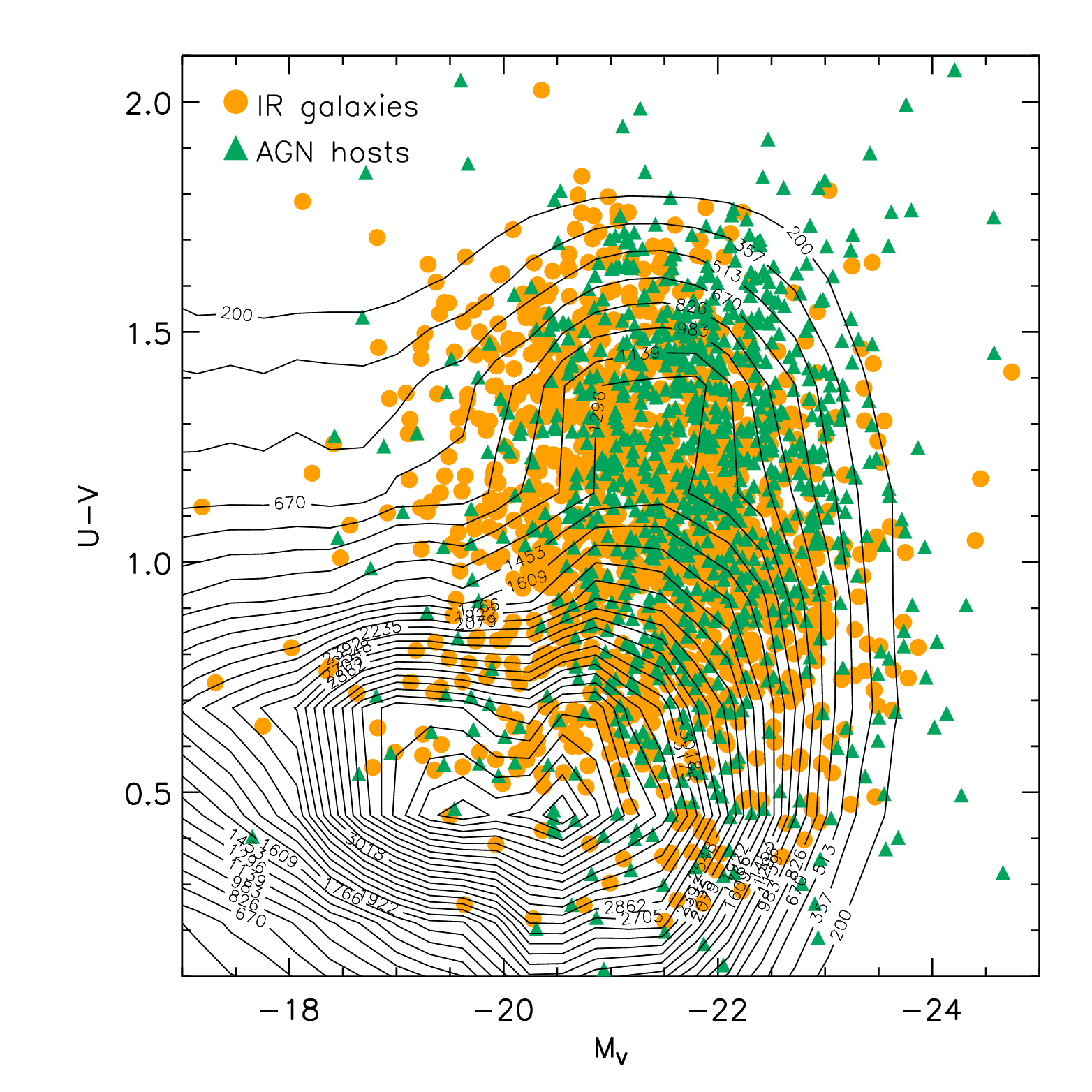,width=0.99\linewidth,clip=} \\
 \epsfig{file=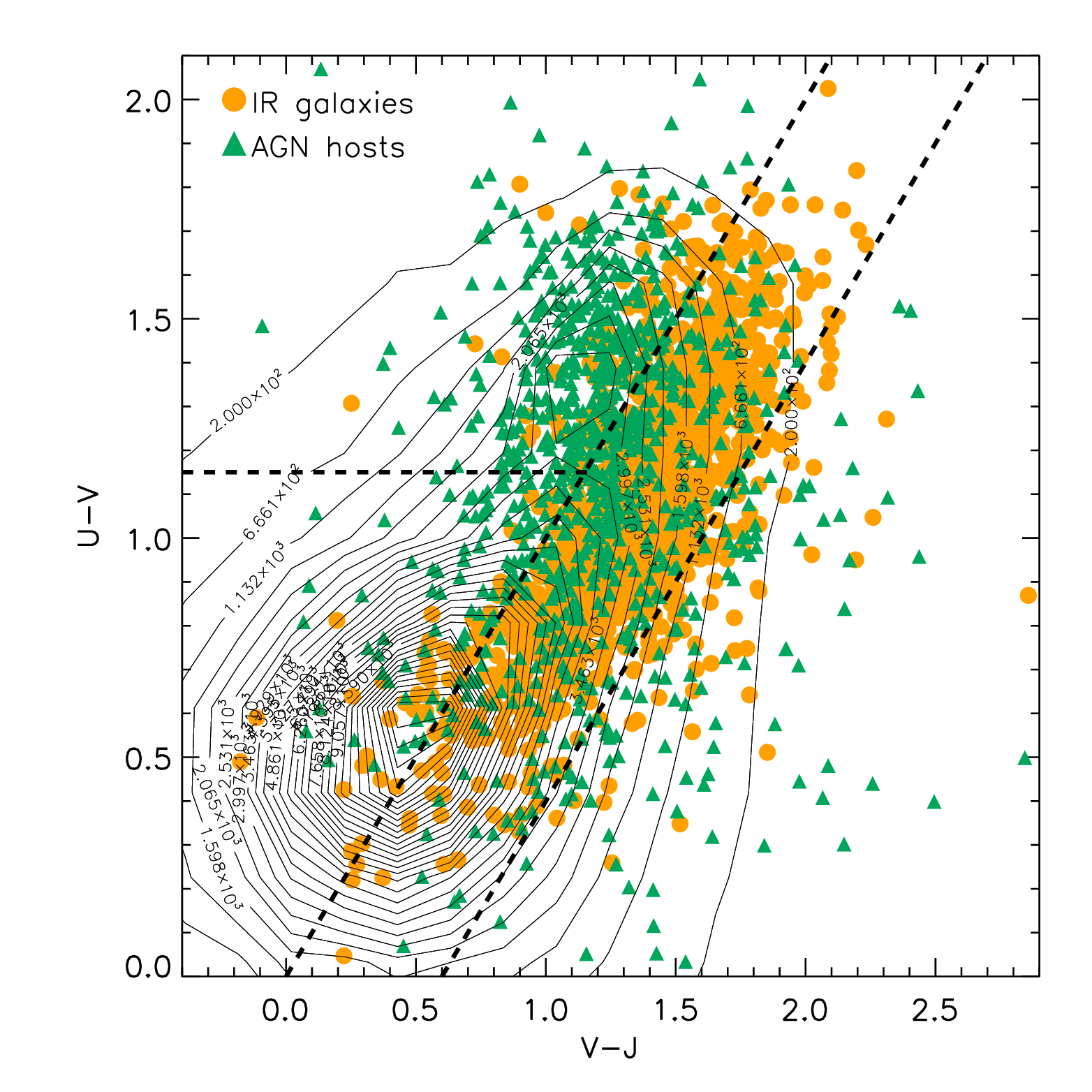,width=0.99\linewidth,clip=} \\
\end{tabular}
     \caption{\textit{Top panel}: The rest-frame $U-V$ vs $M_{\rm V}$ distribution for the IR galaxy sample (orange circles)
        and AGN hosts (green triangles). The contours represent the
        number of optically-selected COSMOS sources in bins of 0.3 mag
        in $M_{\rm V}$ and 0.2 mag in $U-V$. There are 30 contours drawn in
        equal steps starting at 200 and ending at 4740 sources. The `green
        valley' is around $U-V \sim$1.1 (see also Kartaltepe et
        al. 2010b). \textit{Lower panel}: The rest-frame $U-V$ vs $V-J$ distribution for the IR galaxy sample (orange circles)
        and AGN hosts (green triangles). The contours represent the
        number of optically-selected COSMOS sources in bins of 0.2 mag
        in both axes. There are 30 contours drawn in equal steps
        starting at 200 and ending at 13,717 sources. The dashed lines roughly outline the region traditionally
      occupied by age-reddened galaxies (top left quadrant) and region of parameter space occupied
      by the IR-galaxy sample, the `dusty galaxy sequence' (parallel lines). }
         \label{fig:CMD_CCD}
   \end{figure}

\subsection{The sample}
We exclude 26 per cent of the optical sample, as they have a flag
which indicates that they are near a bright source, such as a
star, and so their photometry might be contaminated. We also exclude
from subsequent analysis the $<$6 per
cent of sources which do not have available redshifts. Finally, we cut the 3 datasets (optical, infrared and X-ray)
down to the same coverage
area, leaving us with 304,627 sources in the optically-selected
sample. Out of those, 1274 are detected at 70$\mu$m with
$f_{70}>$7.5mJy ($>$5$\sigma$), hereafter the IR-galaxy sample, and
1394 have $L_{\rm X}>$10$^{42}$\,erg\,s$^{-1}$ in
either the soft or the hard-band, which we take to be our AGN sample. Note that the optical rest-frame magnitudes for the X-ray AGN come from Salvato et al. (2009) and for all other sources from
Ilbert et al. (2009). 

About half of the AGN have
spectroscopic redshifts and the remaining have good quality photometric redshifts from Salvato et
al. (2009\nocite{Salvato09}; 2011\nocite{Salvato11}). 27 per cent of
the AGN are spectroscopically classified as broad line (BLAGN; line FWHM
$>$\,2000\,km\,s$^{-1}$; Brusa et al. 2010). 
For the work presented here, we remove BLAGN, because the galaxy photometry is significantly
contaminated by the active nucleus. We also exclude AGN which do not
have spectroscopic redshifts, but which are classified as type-\Rmnum{1} through
SED fitting to the optical photometry (see Salvato et al. 2009; Bongiorno et al. 2012). This
leaves a total of 856 AGN hosts. Hereafter, when referring to AGN, it is implied that these are all type-\Rmnum{2}
and hence the optical photometry is dominated by the host galaxy. 

The rest-frame colour-magnitude ($U-V$ vs $M_{\rm V}$) and
colour-colour ($U-V$ vs $V-J$) distributions of X-ray AGN hosts and the
IR galaxy sample are shown in Fig. \ref{fig:CMD_CCD}, overlaid on the optically selected COSMOS population. 
The COSMOS sample shows a peak at blue
colours ($U-V\sim$0.5; the `blue cloud') and a peak at red colours ($U-V\sim$1.3; the `red sequence'). The
transition region between the two peaks, at green ($U-V$$\sim$1--1.2)
colours, is more sparsely populated; the so-called `green valley' (see also Kartaltepe et
al. 2010b). We observe a remarkable
overlap between the AGN hosts and IR-galaxies as they span
the same range in $U-V$ colour and cluster
at the high $V$-band luminosity end occupied by massive
galaxies. The colours of the infrared sample are a consequence of dust reddening rather than
age-related reddening, more evident in colour-colour space where age-reddened
galaxies are offset from the sequence of IR galaxies (the `dusty
galaxy sequence'), towards bluer
$V-J$ colours; see also Kartaltepe et al. (2010b); Wuyts et al. (2009\nocite{Wuyts09}); Williams et
al. (2009\nocite{Williams09}); Whitaker et
al. (2011\nocite{Whitaker11}). Particularly noteworthy is that AGN
hosts lie in two camps in colour-colour space: some have colours
consistent with the dusty sequence of star-forming galaxies, whereas a large fraction
have bluer $V-J$ colours, offset from the dusty galaxy
sequence. These characteristics are examined in detail in section
\ref{sec:cmd}. 

Out of the 856 type-\Rmnum{2} AGN hosts in our sample, 66 are also 70\,$\mu$m detected. 
In order to ensure that for these sources the 70\,$\mu$m emission is host-galaxy dominated, i.e. linked to
star-formation, the 70$\mu$m to 24\,$\mu$m flux density ratio is used as a discriminator between AGN and
star-formation dominated far-IR emission, as described in Mullaney et
al. (2010\nocite{Mullaney10}). Note that this criterion can be considered reliable up to
z$\sim$1.5, hence we restrict all subsequent analysis to the
0.1$<z<$1.5 redshift range, where the lower redshift
cut serves to remove local, extended sources. AGN with $f_{70}$/$f_{24}$$>$6 are taken to be host galaxy
dominated in the infrared --- 45 sources in total at 0.1$<z<$1.5, hereafter referred
to as the AGN/starburst (SB) hybrids. 
In addition, we restrict the IR galaxy sample to sources with
$f_{70}$/$f_{24}$$>$6 in order to exclude any sources which
potentially host an AGN not detected in the X-rays but dominant in
the mid/far-IR. 
Our final sample consists of 208,512 optically-selected
sources at 0.1$<z<$1.5, of which 1156 are IR-emitters, 634 host AGN
and 45 are AGN/SB hybrids.

Hereafter, we use the terms `IR galaxies', 
`star-forming galaxies (SFGs)' and `starbursts (SBs)' interchangeably
throughout this paper and `IR
emission' refers to infrared emission from the host galaxy linked to star-formation.

  \begin{figure}
   \centering
\epsfig{file=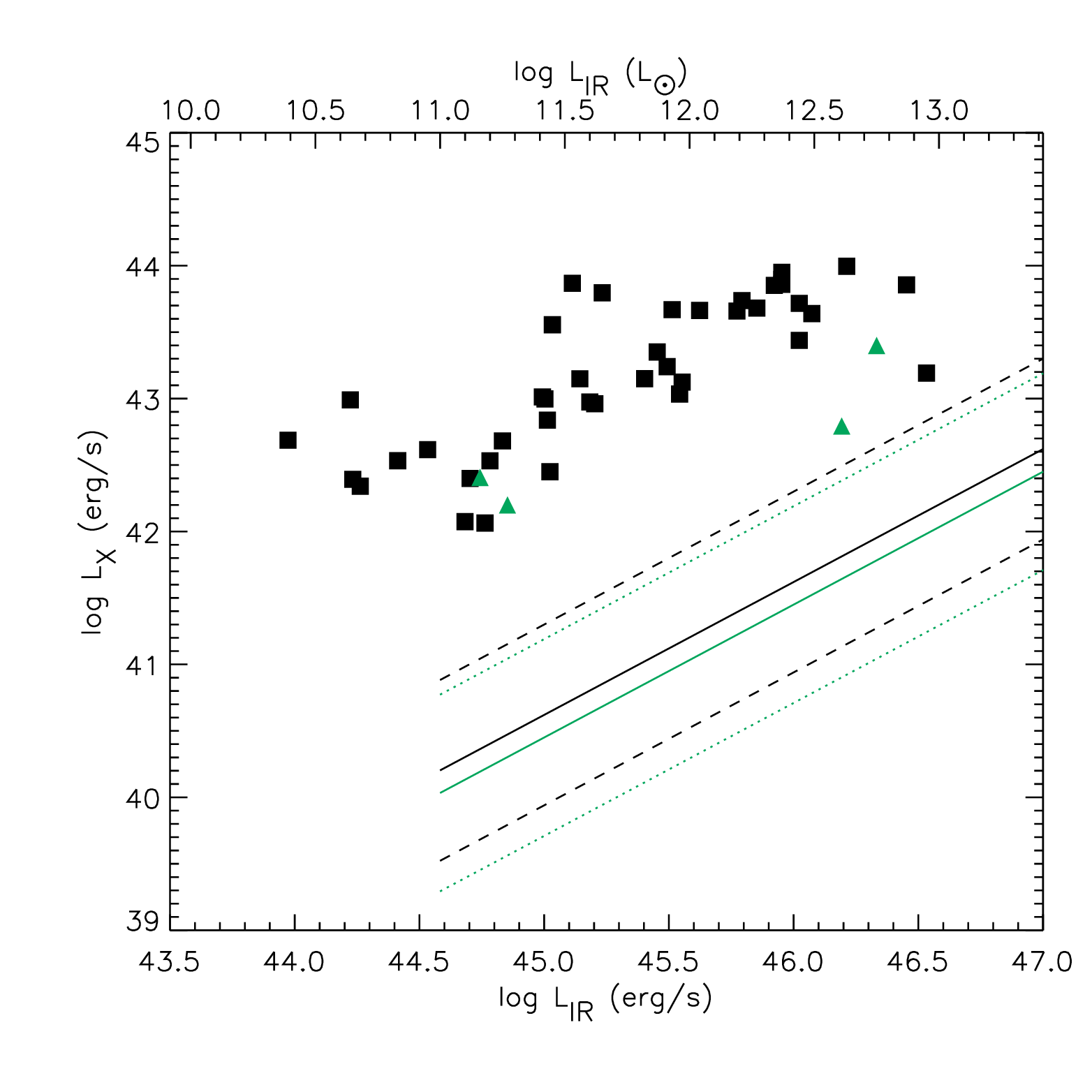,width=0.95\linewidth,clip=} \\
    \caption{X-ray vs total infrared luminosity of the 45 AGN/SB hybrids in
      our final sample. When a source does not have a hard band
      detection (shown as a black square) then
     we plot its soft X-ray luminosity (shown as a green triangle). The solid
     lines are the X-ray/IR correlations for $L_{\rm
       IR}>10^{11}\,L_{\odot}$ galaxies (LIRGs and ULIRGs) which
     are star-formation dominated in both the X-rays and the
     infrared (taken from Symeonidis et al. 2011b). The black
     solid and dashed lines are the \textit{hard} X-ray-IR
       correlation (log\,$L_{\rm HX}$=log\,$L_{\rm
        IR}$-4.38) and $\pm$2$\sigma$ boundaries ($\pm$0.68\,dex). The green solid and dotted lines
are the \textit{soft} X-ray-IR correlation (log\,$L_{\rm SX}$=log\,$L_{\rm
        IR}$-4.55) and $\pm$2$\sigma$ boundaries ($\pm$0.74\,dex). Note that the
      AGN/SB hybrids are offset from the relations for
      SFGs by at least 2$\sigma$, indicating that their X-ray emission can
      be assumed to be entirely AGN dominated. On the other hand their
    infrared emission is star-formation dominated (see main text). }
         \label{fig:xray_ir}
   \end{figure}

  \begin{figure}
   \centering
\epsfig{file=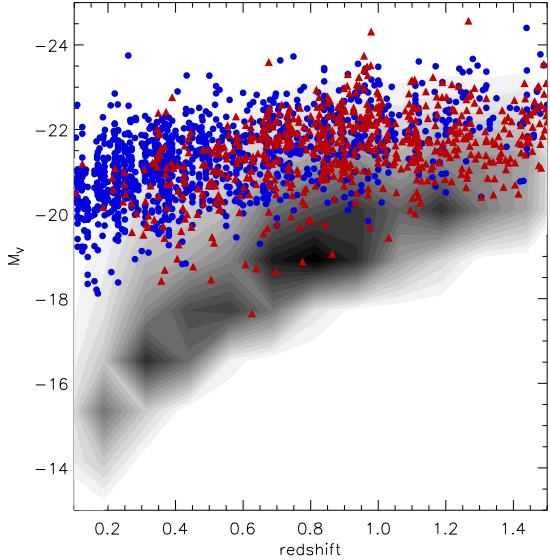,width=0.95\linewidth,clip=} \\
    \caption{V-band absolute magnitude versus redshift for the samples
      used in this work --- the
      optical population (grey filled-in contours), the AGN hosts (red triangles)
    and the IR galaxies (blue circles).}
         \label{fig:MV_z}
   \end{figure}

  \begin{figure}
   \centering
\epsfig{file=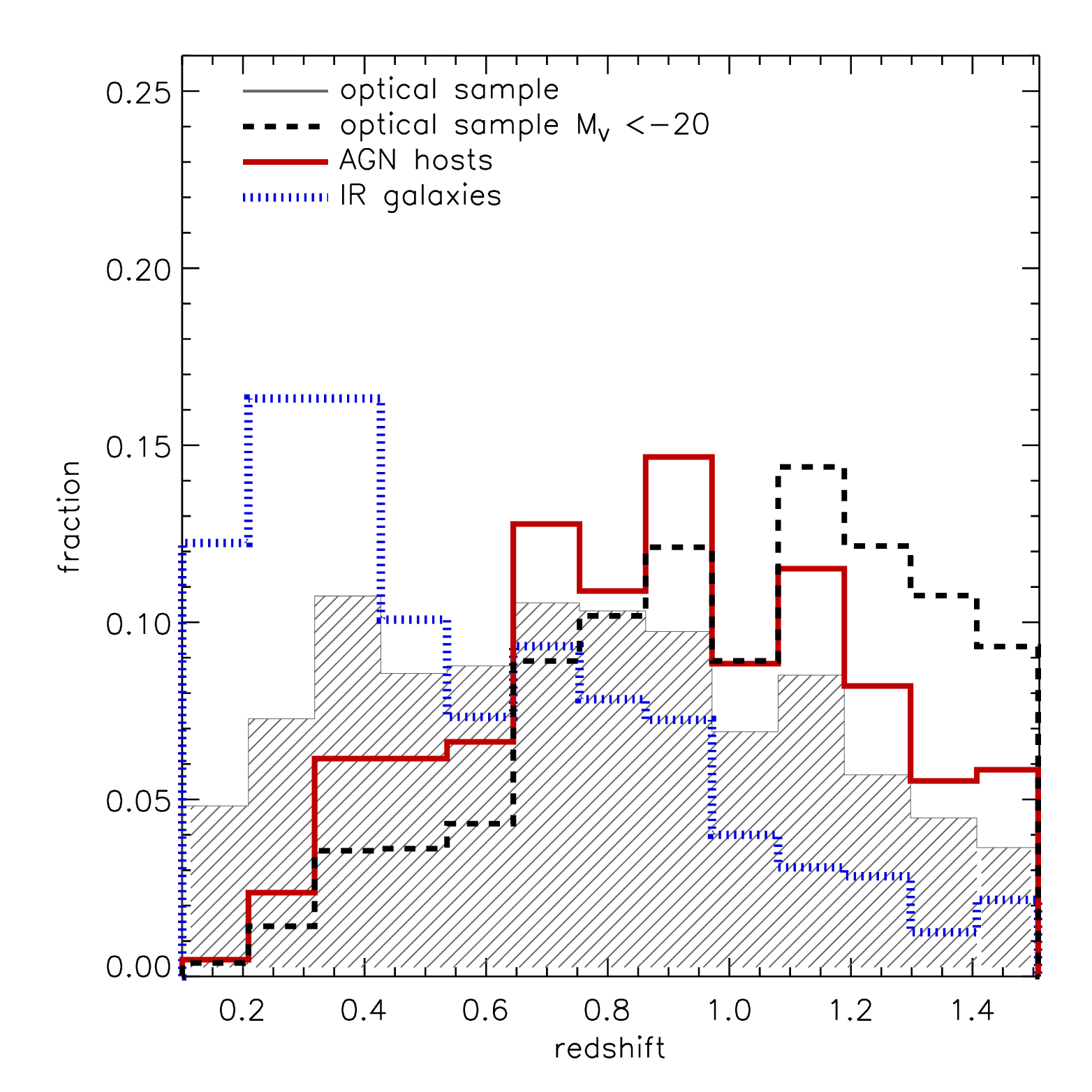,width=0.95\linewidth,clip=} \\
    \caption{The redshift distributions of the samples used in this
      work. Each histogram is normalised to the total number of
      sources in the corresponding sample. Note that although the underlying optically-selected
      population has a relatively flat distribution, restricting
      it to the most optically bright sources ($M_{\rm V}>-20$) 
      skews it towards higher redshifts. }
         \label{fig:z_dist}
   \end{figure}

\subsection{Sample properties}

Fig. \ref{fig:xray_ir} shows the X-ray versus total infrared luminosity
of the 45 hybrid sources, against the X-ray-Infrared correlations for star-forming
galaxies taken from Symeonidis et
al. (2011b)\nocite{Symeonidis11b} for $L_{\rm IR}>10^{11}\,L_{\odot}$
sources (luminous and ultraluminous infrared galaxies, LIRGs and ULIRGs). To calculate X-ray luminosities,
the soft and hard fluxes are \textit{K}-corrected using a photon index
$\Gamma$=1.9, whereas the total infrared luminosities are taken from
Kartaltepe et al. (2010a). We note that the X-ray luminosities of the hybrid
sources are offset by at least 2$\sigma$ from the mean star-formation relations
in the hard and soft X-ray bands. At lower infrared luminosities ($L_{\rm IR}<10^{11}$\,L$_{\odot}$)
there is some evidence that these relations might have a flatter slope (e.g. Lehmer et
al. 2010\nocite{Lehmer10}), nevertheless the $L_{\rm IR}<10^{11}$\,L$_{\odot}$ hybrids in
our sample would still remain well above them. As a result, X-ray emission
from all AGN/SB hybrids can be considered entirely AGN-dominated. In fact, in Symeonidis et
al. (2010)\nocite{Symeonidis10} we showed that a 200\,ks X-ray survey
is largely insensitive to X-ray emission from star-formation, and in the hard band even
a 2Ms survey is insensitive to X-ray emission from star-formation
(Symeonidis et al. 2011b), suggesting that our AGN sample is not contaminated by sources where
star-formation substantially contributes or dominates the X-ray emission. 

Fig. \ref{fig:MV_z} shows absolute V-band magnitude as a
function of redshift in the redshift range of interest (0.1$<z<$1.5), for the optically selected sample, the AGN hosts
and IR galaxies. The AGN hosts and IR galaxies are more optically luminous than the average
optically-selected galaxy at each redshift, suggesting that they are also more massive as optical
luminosity correlates with stellar mass (e.g. Shapley et al. 2001,
Savaglio et al. 2005). Fig. \ref{fig:z_dist} shows the redshift distribution of the 3 galaxy types. We note that
the AGN hosts peak at higher redshift than the IR galaxies, whereas
the general population of optically-selected galaxies have a
relatively flat distribution. Restricting the latter to the brightest
sources with $M_{\rm V}<$-20, which is the range in $M_{\rm V}$ probed
by the AGN and IR galaxies (see Fig. \ref{fig:MV_z}), we find that the
redshift distribution of the optical sample now peaks at higher
redshifts, similar to the AGN hosts.

\section {Results and Analysis}
\label{sec:cmd}

\subsection{The distribution of AGN and IR galaxies in
  colour-magnitude and colour-colour space}

\begin{figure}
   \centering
   \begin{tabular}{c}
   \epsfig{file=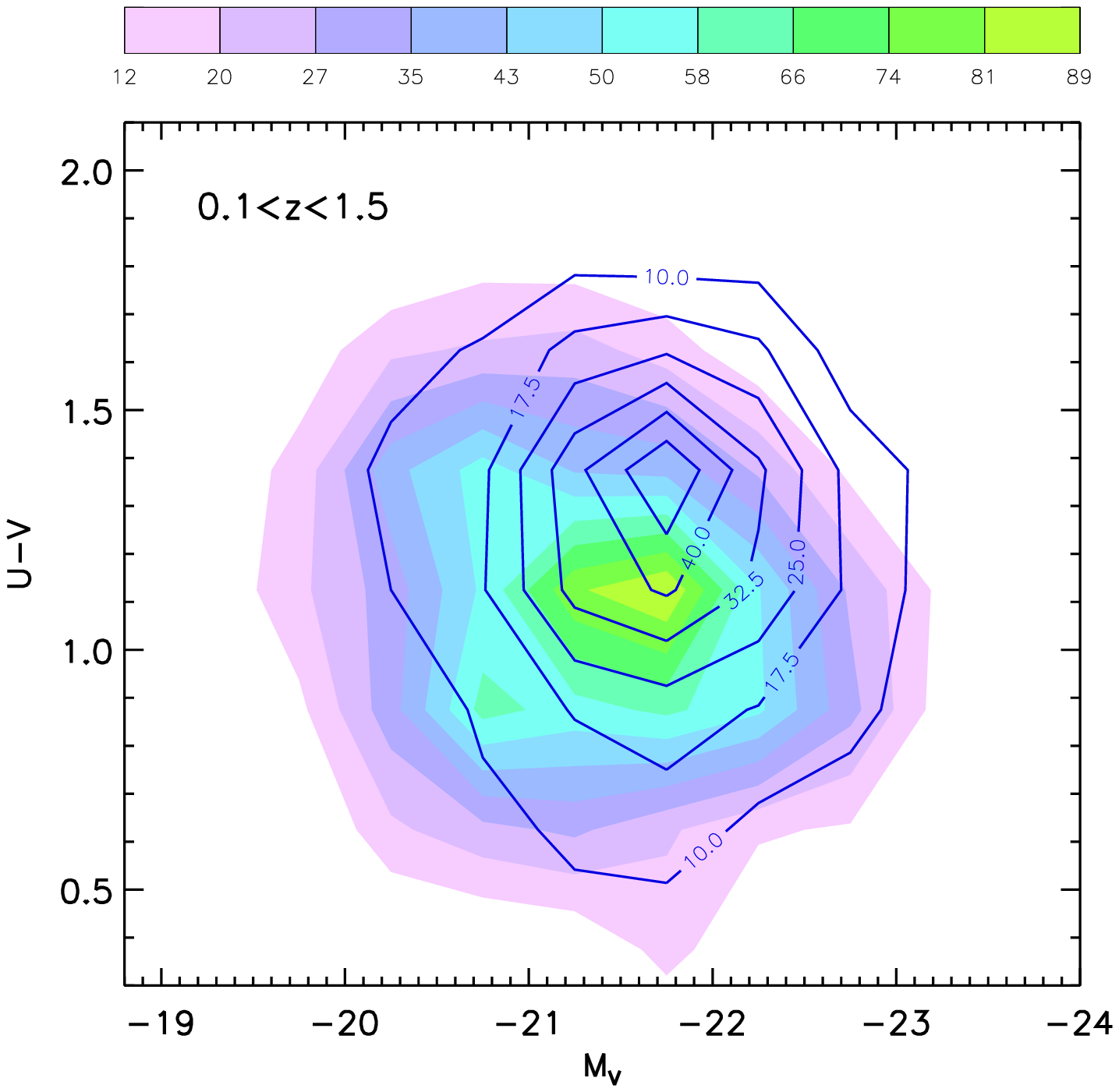,width=0.38\textwidth}\\
   \epsfig{file=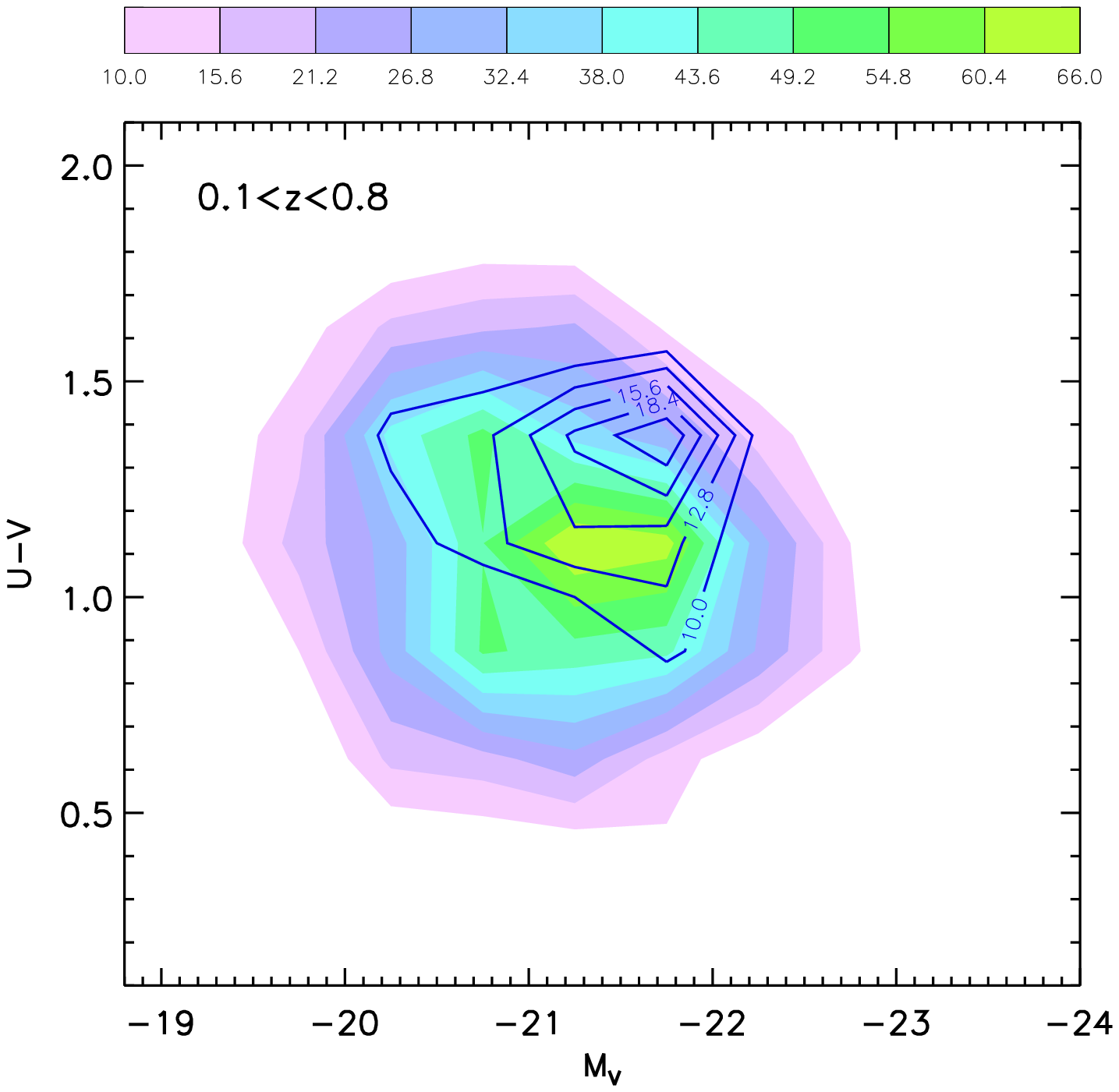,width=0.38\textwidth}\\
   \epsfig{file=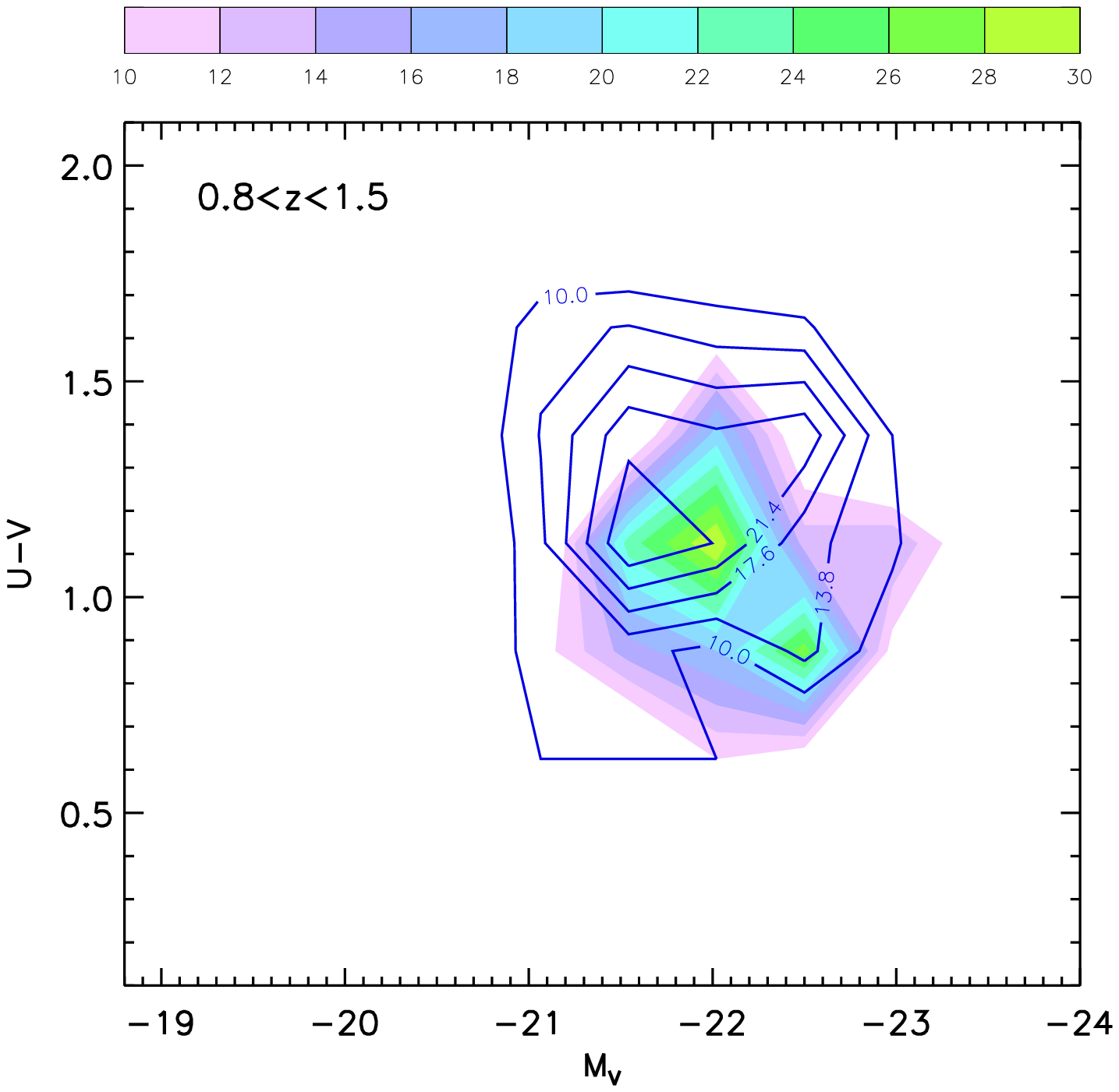,width=0.38\textwidth}\\
  \end{tabular}
     \caption{Rest-frame $U-V$ vs $M_{\rm V}$ for the IR galaxy sample (filled contours)
        and AGN host galaxies (open blue contours); the 3 panels
        correspond to different redshift ranges. The bins are 0.5 and
        0.25 in size in the x and y directions respectively and the contours represent the number of objects. 5 contour levels are drawn for
     the AGN and 10 contour levels for the IR-galaxies (see colour bar), starting from
     $n$ number of objects, where $n$ is either $0.01\,N$ (where $N$ is
     the total number of sources) or 10, whichever is greater. }
         \label{fig:CMD1}
   \end{figure}

\begin{figure}
   \centering
   \begin{tabular}{c}
   \epsfig{file=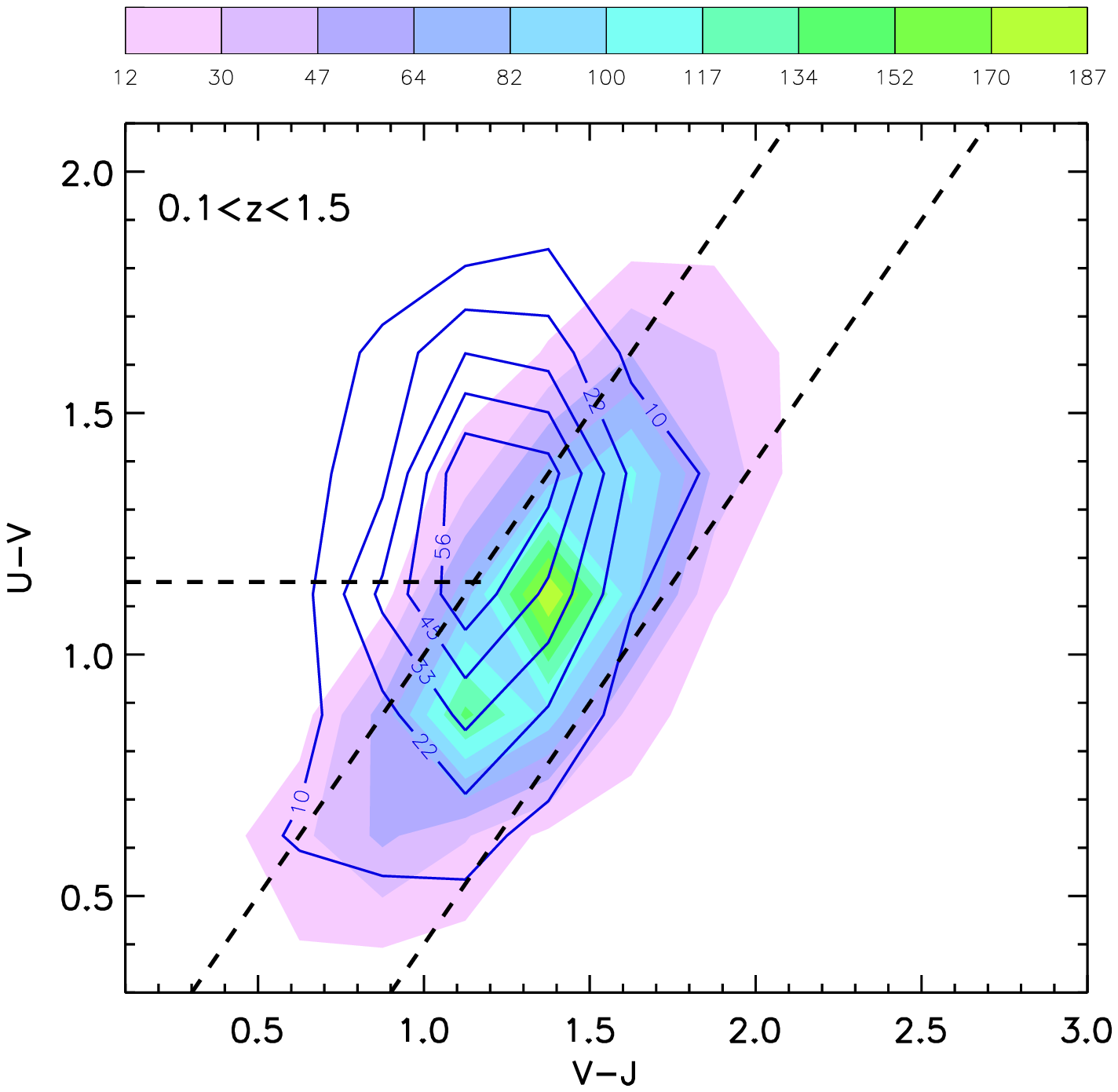,width=0.38\textwidth}\\
   \epsfig{file=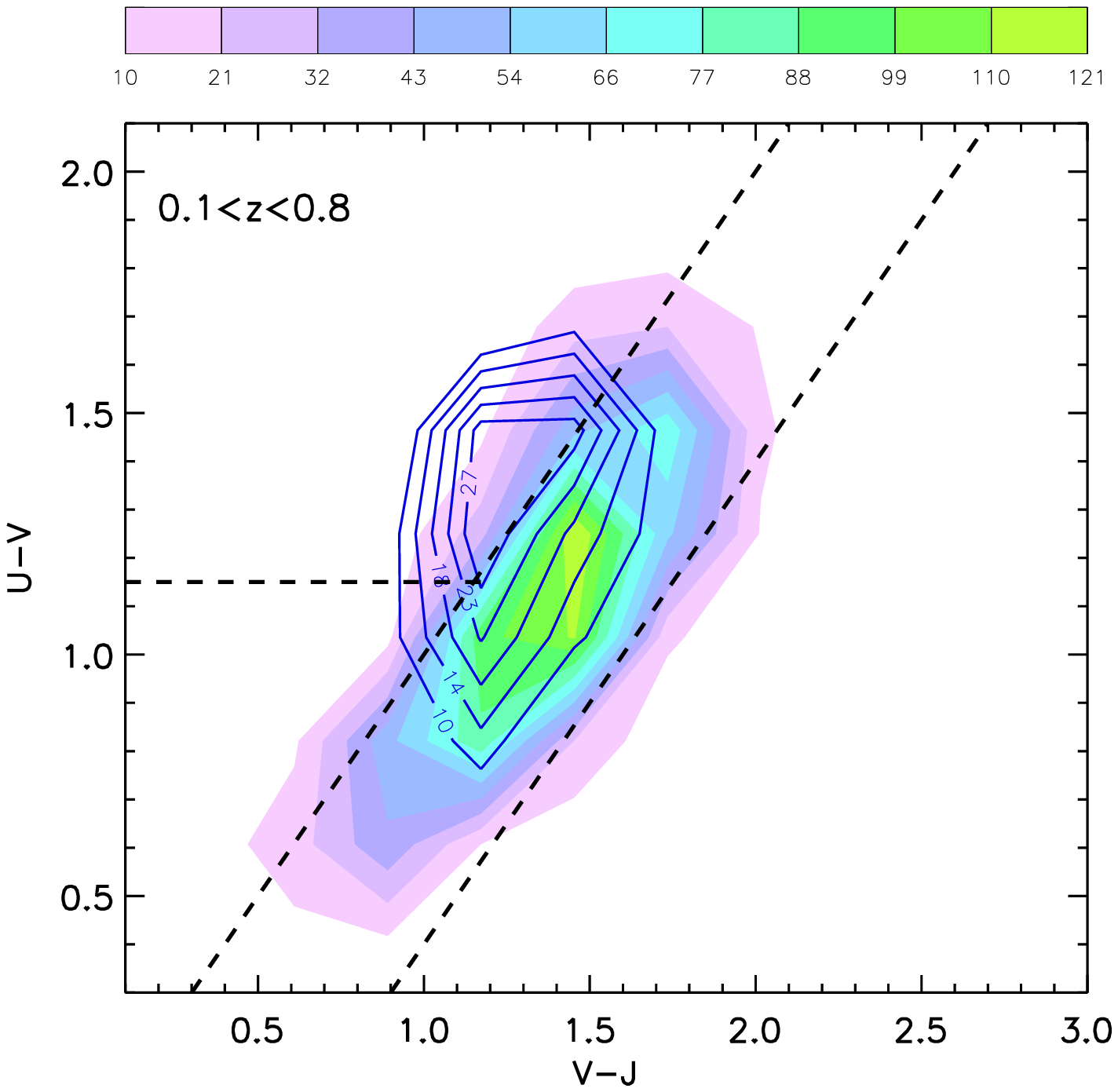,width=0.38\textwidth}\\
   \epsfig{file=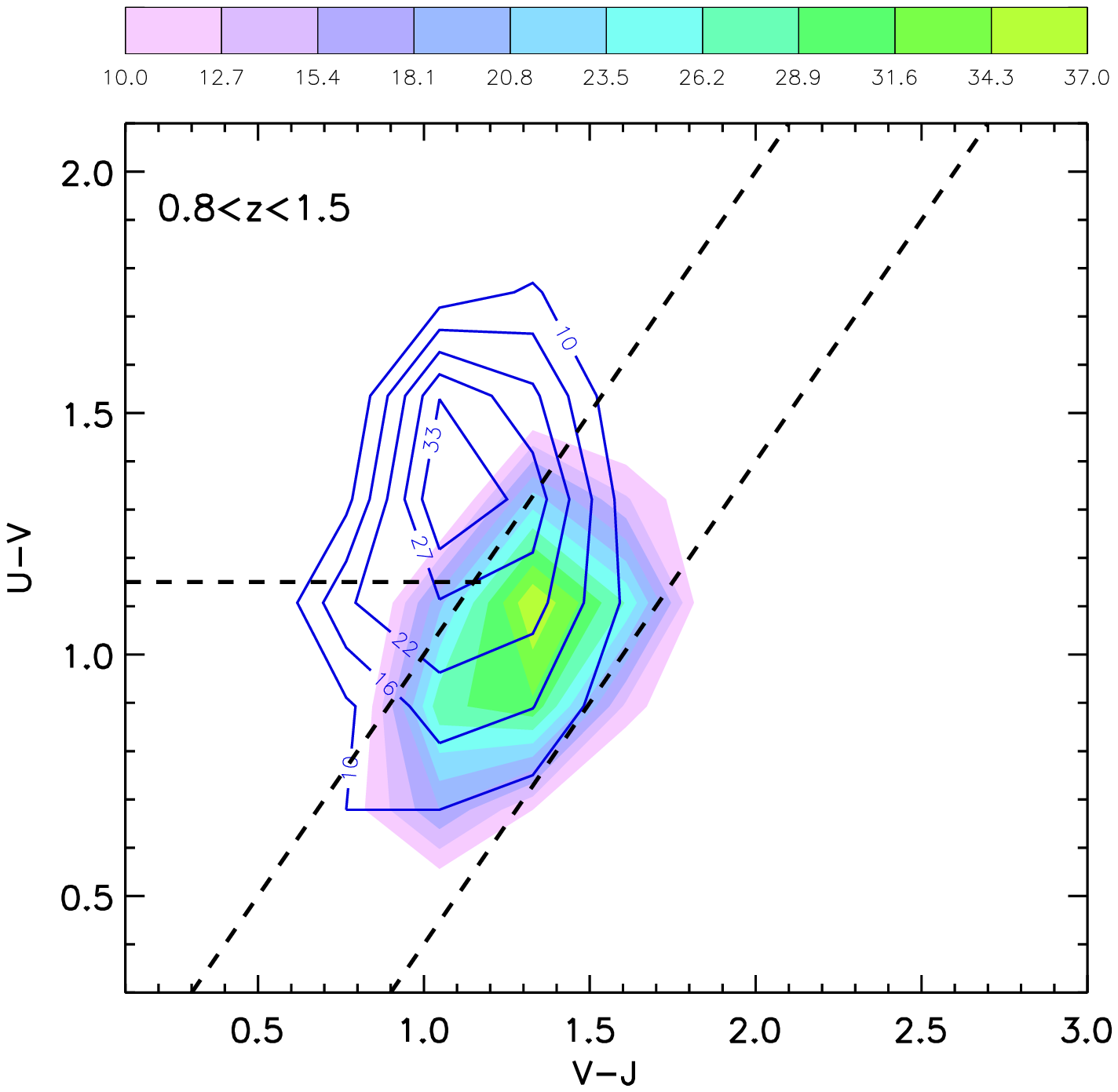,width=0.38\textwidth}\\
  \end{tabular}
     \caption{Rest-frame $U-V$ vs $V-J$ for the IR galaxy sample (filled contours)
        and AGN host galaxies (open blue contours); the 3 panels
        correspond to different redshift ranges. The dotted lines
        are the same as in Fig. \ref{fig:CMD_CCD}. The bins are 0.25
        in both axes and the contours represent the number of objects. 5 contour levels are drawn for
     the AGN and 10 contour levels for the IR-galaxies (see colour bar), starting from
     $n$ number of objects, where $n$ is either $0.01\,N$ (where $N$ is
     the total number of sources) or 10, whichever is greater. }
         \label{fig:CCD1}
   \end{figure}

In Figs \ref{fig:CMD1} and \ref{fig:CCD1} we show the colour-magnitude
and colour-colour distributions for the AGN hosts and IR
galaxies. They are binned and displayed as contours and further split into two
redshift bins ($0.1<z<0.8$ and $0.8<z<1.5$). The overlap noted in Fig. \ref{fig:CMD_CCD} is now seen more
clearly, however differences between the two
populations also begin to emerge. We note that the peaks of the two
distributions are well aligned in $M_{\rm V}$ but not in $U-V$, with AGN
hosts peaking at redder colours. Moreover, many AGN hosts show bluer
$V-J$ and redder $U-V$ colours consistent with those of red-sequence
galaxies, offset to the left of the dusty galaxy sequence. These observations give weight to the notion that AGN host galaxies are not a
uniform population. They also suggest that some AGN hosts have colours indicative of star-formation and
dust-extinction, whereas others have colours consistent with an
overall slowing down or termination of star-formation.

  \begin{figure}
   \centering
   \begin{tabular}{c}
   \epsfig{file=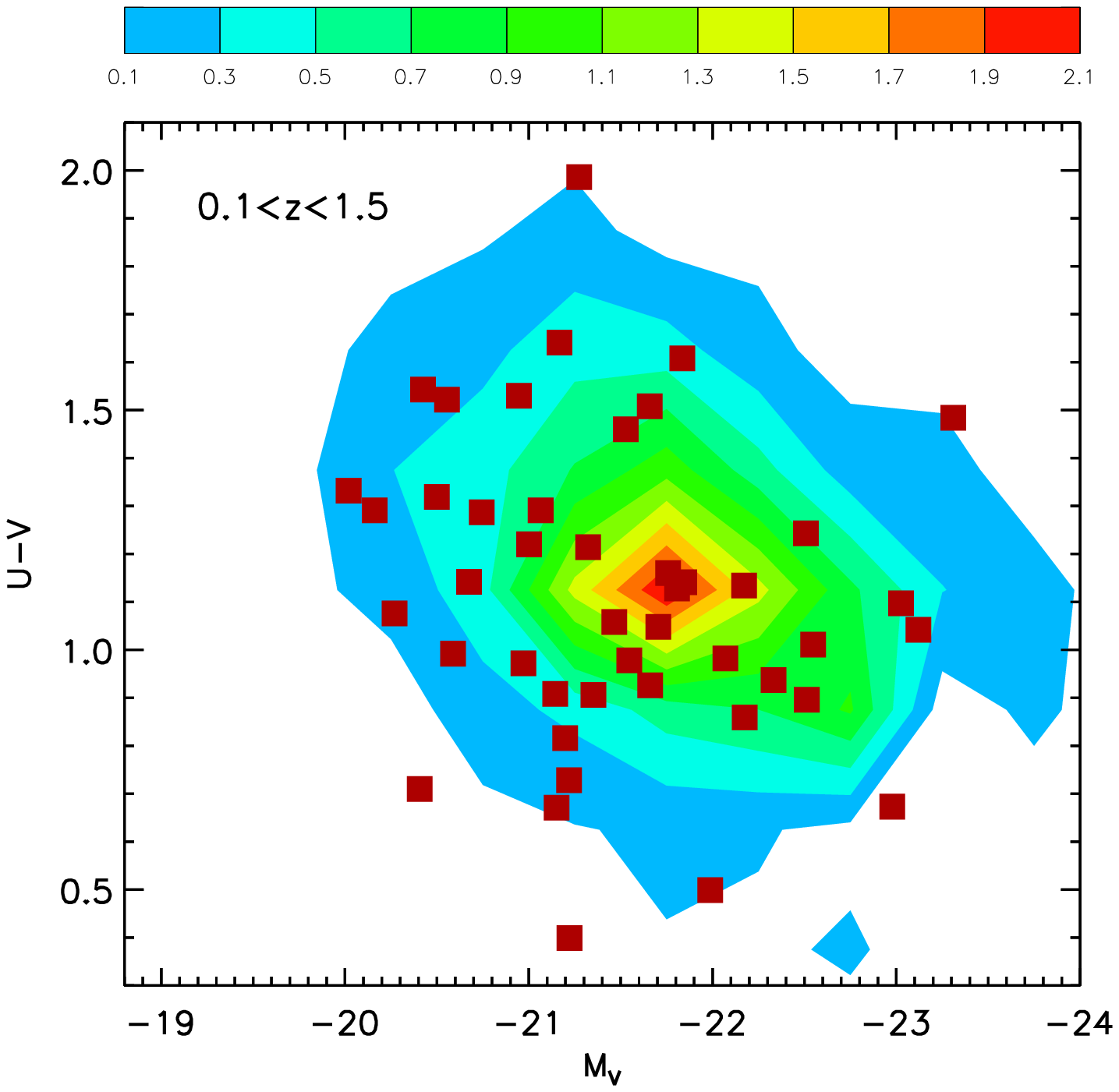,width=0.38\textwidth}\\
   \epsfig{file=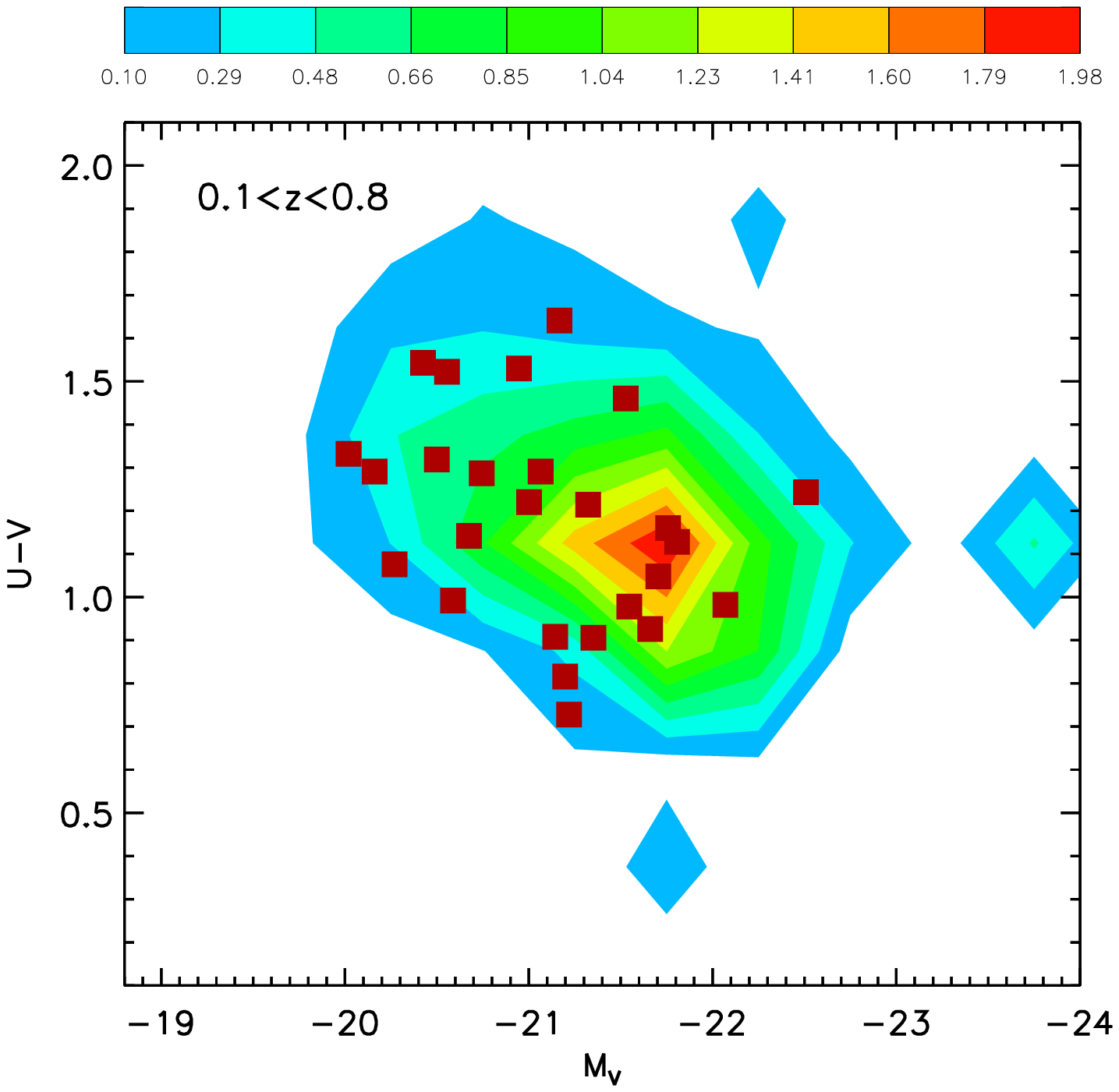,width=0.38\textwidth}\\
   \epsfig{file=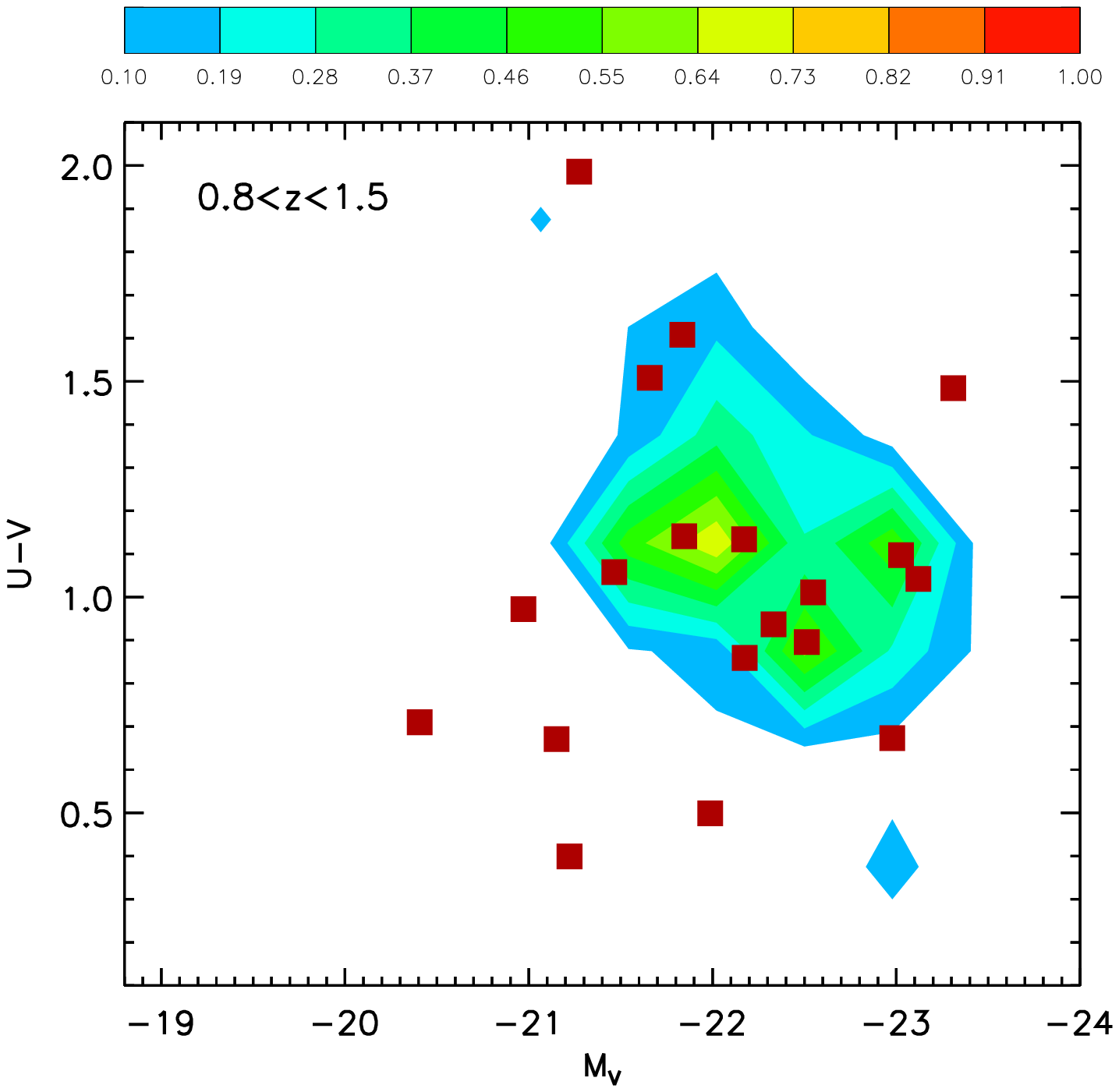,width=0.38\textwidth}\\
  \end{tabular}
     \caption{ The colour-magnitude distribution expected for AGN/SB
       hybrids and computed under the assumption
       that AGN and star-formation are independent events. The total
       number of expected hybrids ($N_{\rm exp}$=15.9$\pm$1 for the
       top panel) is calculated by summing the number of hybrids
       in each bin. In all panels, the bins are 0.5 and 0.25 in size in the x and y
       directions respectively and contours correspond to the number
       of sources (see colourbar). Red squares indicate the observed
       distribution of AGN/SB hybrids. The 3 panels correspond to
       different redshift ranges. }
         \label{fig:CMD2}
   \end{figure}

  \begin{figure}
   \centering
   \begin{tabular}{c}
   \epsfig{file=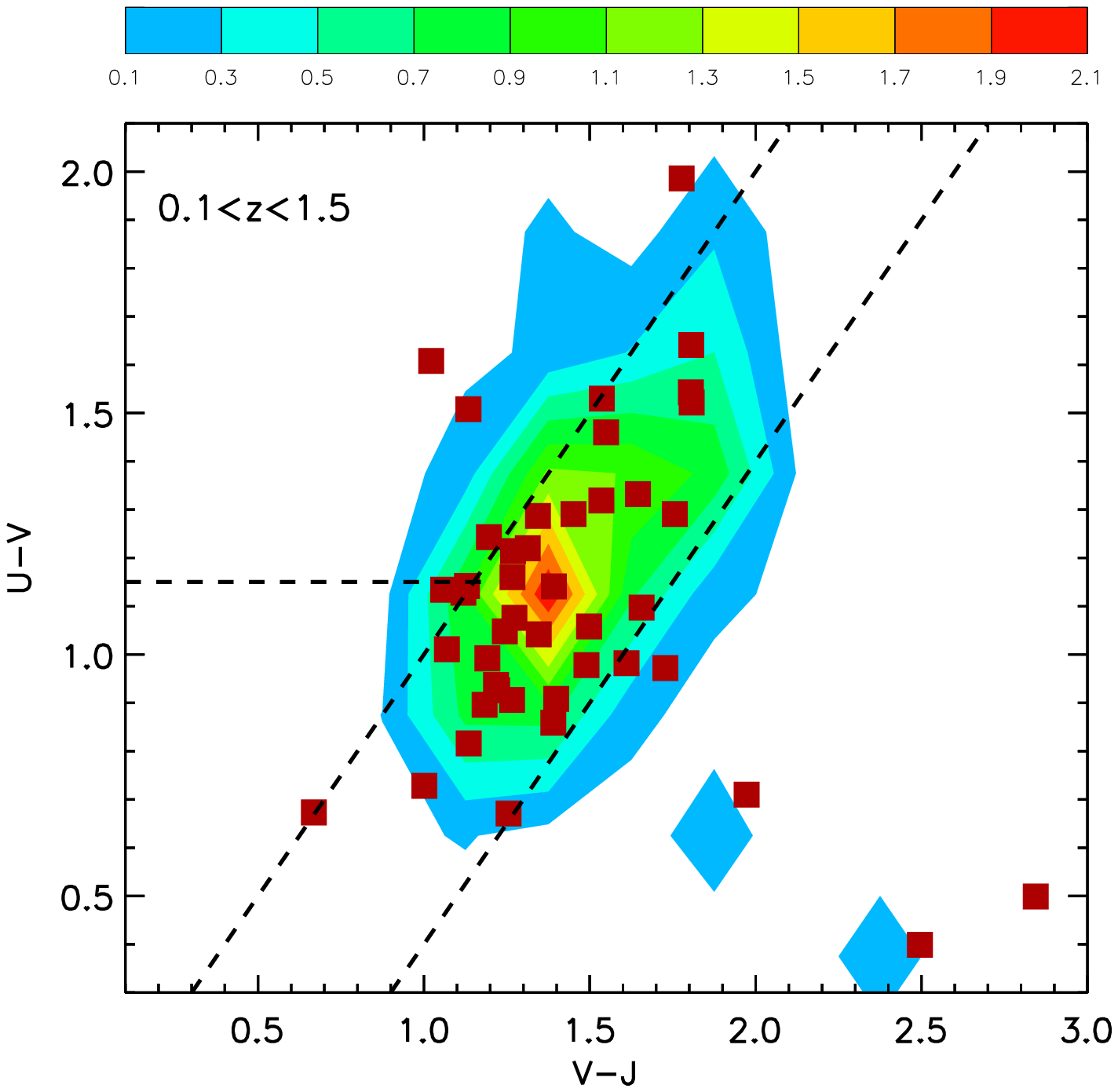,width=0.38\textwidth}\\
   \epsfig{file=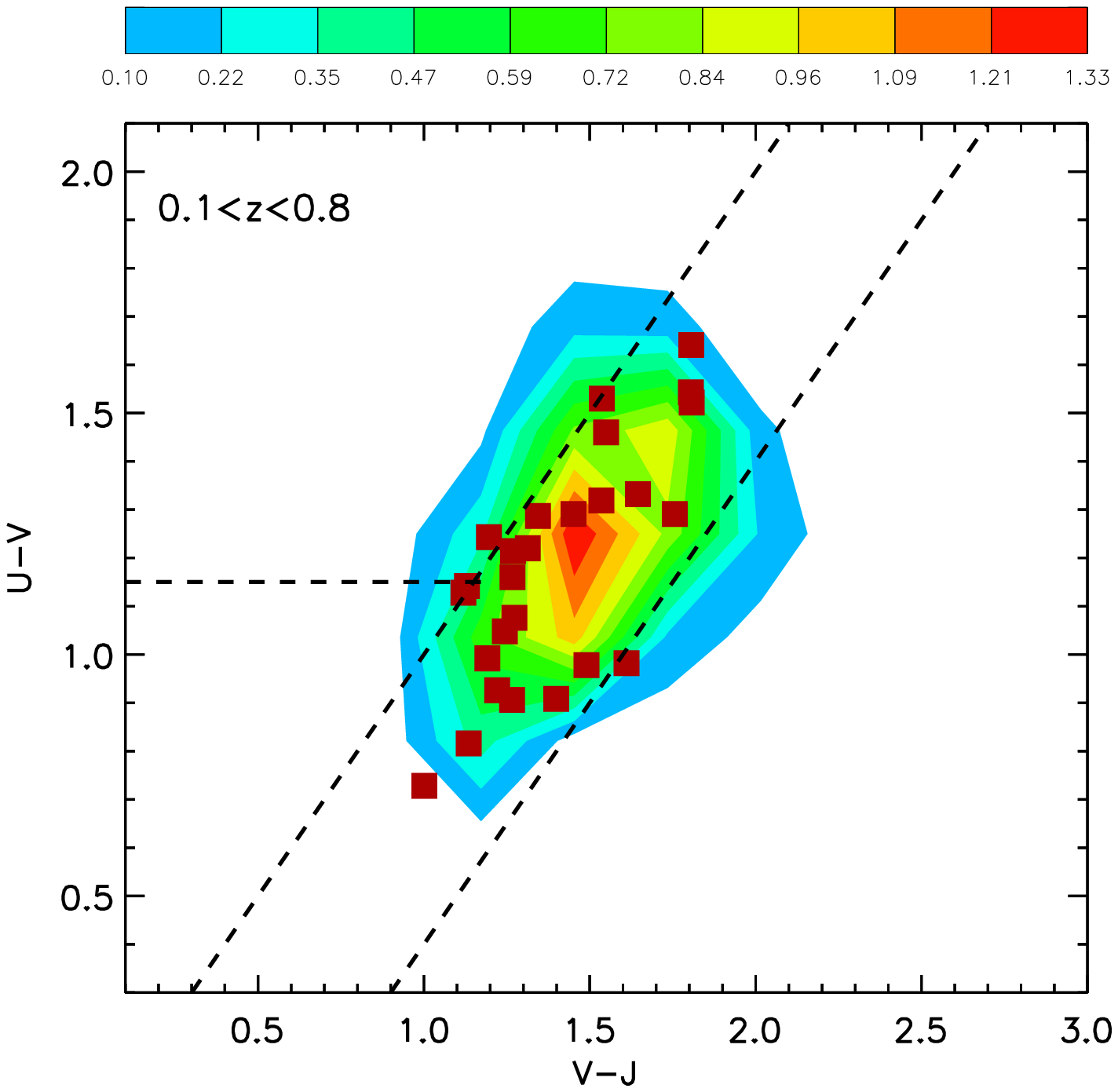,width=0.38\textwidth}\\
   \epsfig{file=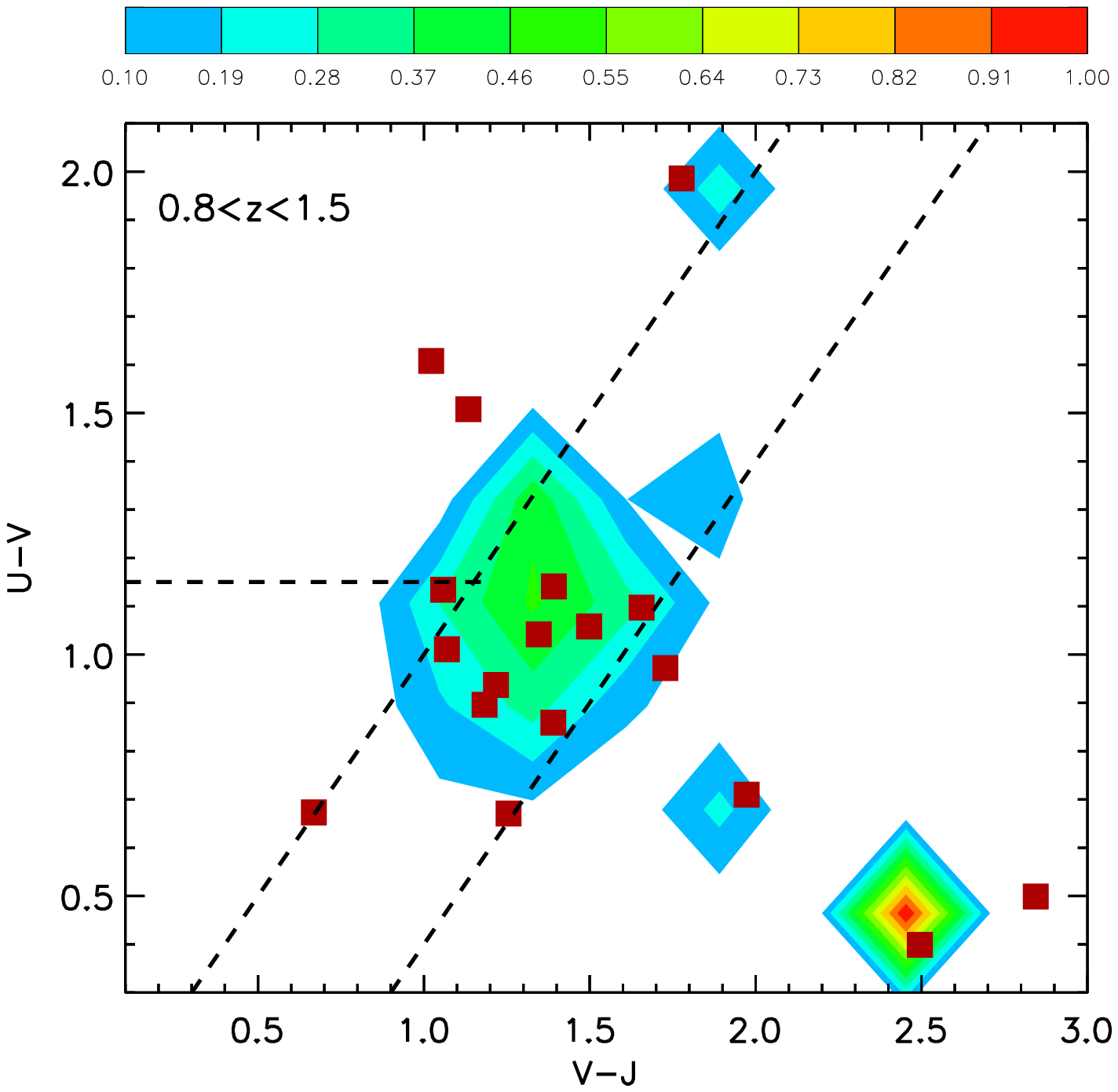,width=0.38\textwidth}\\
  \end{tabular}
     \caption{The colour-colour distribution expected for AGN/SB
       hybrids and computed under the assumption
       that AGN and star-formation are independent events. The total
       number of expected hybrids ($N_{\rm exp}$=12$\pm$0.8 for the
       top panel) is calculated by summing the number of hybrids
       in each bin. In all panels, the bins are 0.25 and 0.25 in size in the x and y
       directions respectively and contours correspond to the number
       of sources (see colourbar). Red squares indicate the observed
       distribution of AGN/SB hybrids. The 3 panels correspond to
       different redshift ranges. }
         \label{fig:CCD2}
   \end{figure}

\begin{table*}
\centering
\caption{Table showing the results of our experiment which compared
  the observed number of AGN/SB hybrids ($N_{\rm obs}$=45$\pm$6.7) to the
  expected number of AGN/SB hybrids ($N_{\rm exp}$) computed by assuming that AGN and SF
  are independent events in any given galaxy (see Section \ref{sec:cmd1}). In the first column we
  show the redshift binning. `none' indicates that the experiment is performed over
the whole redshift range ($0.1<z<1.5$) and so the samples are only
binned in colour and magnitude, whereas for 2 and 4 bins, the samples
are also binned in redshift. The table examines whether $N_{\rm exp}$ is
significantly different to $N_{\rm obs}$, separately
for the colour-magnitude and colour-colour parameter space. $N_{\rm exp}$ (columns 2 and
5) is between 2 and 4 times lower than $N_{\rm obs}$, indicating that
type-\Rmnum{2} AGN are on average 3 times more likely to reside in dusty
SFGs than would be expected serendipitously. Our calculations show
that this result is significant at least at the 3\,$\sigma$ level
and on average at the 4\,$\sigma$ level (columns 4 and 7). The
significance is calculated by dividing the value of [$N_{\rm obs} - N_{\rm
  exp}$] by its uncertainty (columns 3 and 6).}
\begin{tabular}{|c|c|c|c|c|c|c|}
\hline 
&\multicolumn{3}{|c|}{colour-magnitude} & \multicolumn{3}{|c|}{colour-colour} \\ 
\hline
redshift binning & $N_{\rm exp}$ & $N_{\rm obs} - N_{\rm exp}$ & significance&
$N_{\rm exp}$ & $N_{\rm obs} - N_{\rm exp}$ & significance \\ 
\hline
none & 15.9$\pm$1&29.1$\pm$6.8 &4.3\,$\sigma$ & 12$\pm$0.8&33$\pm$6.8 &4.9\,$\sigma$\\
2 bins& 18.7$\pm$1.4&  26.3$\pm$6.8&3.8\,$\sigma$&12$\pm$0.9 &33$\pm$6.8 &4.9\,$\sigma$\\
4 bins& 18.6$\pm$1.5& 26.4$\pm$6.9 & 3.8\,$\sigma$& 11.6$\pm$1& 33.4$\pm$6.8&4.9\,$\sigma$\\
\hline
\end{tabular}
\label{table:hybrids}
\end{table*}

\subsection{The incidence of AGN hosted by IR galaxies}
\label{sec:cmd1}
The high incidence of AGN and IR galaxies in similar parts of colour-magnitude
and colour-colour space raises an important question: `Is it simply a
coincidence?'
To address this point, we start by examining what we would expect if the
answer were `yes' and make the following
probability argument: If AGN incidence and IR emission were
independent, unrelated events in a
galaxy of a given location in colour-magnitude or colour-colour space, what characteristics
should the distributions of those hybrid AGN/SB
sources have? To answer this question, we compute the expected number
($N_{\rm exp}$)  and
distribution of hybrid AGN/SB sources, under
the assumption that IR-emission and AGN incidence in a given galaxy
are independent events. We take binned colour-magnitude (bin size of 0.5
in $M_{\rm V}$ and 0.25 in $U-V$; Fig. \ref{fig:CMD1} top panel) and
colour-colour distributions (bin size of 0.25
in $V-J$ and 0.25 in $U-V$; Fig. \ref{fig:CCD1} top panel), allowing us to
probe sources with similar optical properties (in each bin).
For each bin, we count the number of optically selected sources, IR
galaxies and AGN hosts and calculate the probability that an optically-selected source is
both star-forming and hosts an AGN:
\begin{equation}
\frac{n_{i, \rm AGN}}{n_{i, \rm opt}} \times \frac{n_{i, \rm
    IR}}{n_{i, \rm opt}}= P_i ({\rm AGN}) \times P_i ({\rm IR}) = P_i (\rm
  AGN \cap \rm IR)
\end{equation}
where $n_{i, \rm opt}$, $n_{i, \rm AGN}$ and $n_{i, \rm IR}$ are
the number of optically selected galaxies, AGN and IR-selected
galaxies respectively in a colour-magnitude or colour-colour bin
$i$, and $P_i ({\rm AGN})$ is the probability that
a galaxy hosts an AGN, whereas $P_i ({\rm IR})$ is the 
probability that a galaxy is IR-luminous and hence intensely star-forming.
Subsequently we compute the expected number of galaxies in bin $i$, which are both
IR-luminous and host an AGN:
\begin{equation}
N_{i, \rm exp}=P_i ({\rm AGN} \cap {\rm IR}) \times n_{i, \rm opt}
\end{equation}
Finally, the total number of expected
hybrids $N_{\rm exp}$ is calculated by summing over all bins, i.e.
\begin{equation}
N_{\rm exp} = \sum_{\substack{i}} N_{i, \rm exp}
\end{equation}
For each bin, the error on the expected number, $N_{i, \rm exp}$, is calculated by
computing binomial errors on $P_i (\rm AGN)$ and $P_i (\rm IR)$ at the
68 per cent confidence interval ($\sim$1$\sigma$) and assuming no
error on $n_{i, \rm opt}$. The errors on $N_{i, \rm exp}$ from all
bins are then
  added in quadrature to estimate the uncertainty on $N_{\rm
    exp}$. 
The expected colour-magnitude and colour-colour distributions
(i.e. $N_{i, \rm exp}$) are shown in Figs \ref{fig:CMD2} and \ref{fig:CCD2}, for the whole
redshift range $0.1<z<1.5$ but also split into low ($0.1<z<0.8$) and high ($0.8<z<1.5$) redshift. In both colour-magnitude and colour-colour
diagrams, the peak of the expected distribution lies at $U-V \sim$1.1, indicating that the region where we would
expect the highest number of AGN/SB hybrid sources to exist by chance
is at green $U-V$ colours. 

Note that up to now we have not taken into account the redshift
information when computing $N_{\rm exp}$. As seen in Fig. \ref{fig:z_dist}, IR galaxies peak at
lower redshifts than type-\Rmnum{2} AGN
hosts and the optically selected population in the $M_{\rm V}$ range
of interest. To investigate whether the differences in the redshift
distributions have an impact on $N_{\rm exp}$, we repeat the above process, this time also binning in
redshift --- i.e. bin $i$ is now a
  colour-magnitude-redshift bin or a colour-colour-redshift
  bin. We perform the experiment with 2 redshift bins $0.1<z<0.8$ and
  $0.8<z<1.5$ and subsequently with 4 redshift bins of equal size
  ($0.1<z<0.45$, $0.45<z<0.8$, $0.8<z<1.15$, $1.15<z<1.5$). The computed values
  of $N_{\rm exp}$ and corresponding uncertainties are shown in table
  \ref{table:hybrids}. We see that binning in redshift does not
  substantially change $N_{\rm exp}$ and it is difficult to say whether
  the small changes are due to the different redshift distributions or statistical
  fluctuations which come into play when binning (small number
  statistics). 

Note that $N_{\rm exp}$ is higher in colour-magnitude than
  colour-colour space. This is because the two diagrams
  illustrate different galaxy properties and it is evident that there
  is a larger overlap between type-\Rmnum{2} AGN hosts and IR-galaxies in colour-magnitude space than in
colour-colour space (see top panels of Fig. \ref{fig:CMD1} and
\ref{fig:CCD1}).

\begin{table}
\centering
\caption{Table showing the significance of the K-S test performed in
  order to
  evaluate the differences in the
  colours and magnitudes of the observed hybrids and the expected distribution
  of hybrid sources (see Figs \ref{fig:CMD2} and \ref{fig:CCD2}).}
\begin{tabular}{l|l|l|l|l|}
\hline 
& \multicolumn{4}{|c|}{Significance of K-S test} \\ 
\hline
& \multicolumn{2}{|c|}{colour-magnitude} &
\multicolumn{2}{|c|}{colour-colour} \\ 
\hline
Redshift bin & $M_{\rm V}$& $U-V$& $V-J$ &$U-V$  \\ 
\hline
$0.1<z<1.5$&2.5$\sigma$ & 1.2$\sigma$ &2.4$\sigma$&1.7$\sigma$\\
$0.1<z<0.8$& 2.6$\sigma$ & $<$1$\sigma$ & 2.3$\sigma$&$<$1$\sigma$\\
$0.8<z<1.5$& 1.3$\sigma$ & $<$1$\sigma$ & 1.1$\sigma$ &$<$1$\sigma$\\
\hline
\end{tabular}
\label{table:KS}
\end{table}

We remind the reader that observed number of hybrids ($N_{\rm obs}$) is 45 with a Poisson
  error of $\sqrt{45}$ (=6.7). To examine whether the difference between $N_{\rm
    exp}$ and $N_{\rm obs}$ is significant, we divide [$N_{\rm obs} -
  N_{\rm exp}$] by its corresponding uncertainty (see table \ref{table:hybrids}). We find that in all cases $N_{\rm exp}$
  is significantly lower than $N_{\rm obs}$ by at least
  3\,$\sigma$ and on average by 4\,$\sigma$. The ratio of $N_{\rm
        obs}$ to $N_{\rm exp}$ (see table \ref{table:hybrids})
indicates that type-\Rmnum{2} AGN are on average 3 times more likely to reside in dusty
SFGs than would be expected serendipitously. 
Regarding the question posed earlier, our results imply that the existence of AGN hosts and IR-luminous galaxies in the
same part of colour-magnitude and colour-colour space is \emph{not}
coincidental.
Instead it is likely symptomatic of a link between black hole accretion and
star-formation. We speculate that this is a consequence of the large reservoirs of gas (e.g. Ivison et
al. 2011\nocite{Ivison11}) or merger events (e.g. Kartaltepe et
al. 2010b; 2012\nocite{Kartaltepe12}) that often characterise
IR-luminous galaxies, creating favourable conditions both for the AGN and
star-formation. 

Note that although we identify only a small number of AGN/SB hybrids, we are sampling a small part of the $L_{\rm IR}$-$z$
space, as the COSMOS 70\,$\mu$m survey only picks up the most
IR-luminous systems at each redshift slice, i.e. we can
only detect AGN hosts with high SFRs. Lowering the SFR
threshold, e.g. with a deeper 70\,$\mu$m survey, might allow us to
detect star-formation in a larger fraction of AGN hosts, in agreement with reports from other studies e.g. Brusa et al. (2009\nocite{Brusa09}); Mainieri et
al. (2011\nocite{Mainieri11}); Mullaney et
al. (2012a\nocite{Mullaney12a}), Santini et
al. (2012\nocite{Santini12}). Nevertheless, the nature of this study, which
targets the most luminous systems at each redshift slice, allows us
to conclude that the incidence of luminous type-\Rmnum{2} AGN in strongly
star-forming galaxies is not a serendipitous event.

\subsection{The distribution of AGN/SB hybrids in colour-magnitude and
colour-colour space}

We next examine whether the connection between AGN incidence and
IR-emission identified above, also manifests as a difference between the observed and
expected distributions of AGN/SB hybrids, shown in Figs \ref{fig:CMD2} and \ref{fig:CCD2}. We perform 1D K-S tests on the expected and
observed distributions in $U-V$ colour and V-band magnitude (Fig
\ref{fig:CMD2}) and $U-V$ colour and $V-J$ colour (Fig
\ref{fig:CCD2}); the results are shown in table
\ref{table:KS}. We find that the expected and observed
distributions are marginally different with respect to $M_{\rm V}$
and $V-J$. Looking at Fig \ref{fig:CMD2}, we note that the $V$-band
luminosities of the AGN/SB hybrids (particularly obvious in the low
redshift bin) seem lower than
expected. Similarly, Fig
\ref{fig:CCD2} shows that the AGN/SB hybrids are slightly offset to bluer $V-J$
colours from the peak of the expected distribution (particularly
obvious at low redshift). Although the significance of this result is
marginal, it suggests that the
connection between black hole accretion and star-formation,
established earlier, might have an imprint on the properties of a galaxy when these processes occur
simultaneously. Such hybrid sources appear to have
diminished optical emission relative to IR emission and bluer $V-J$
colours. If the reason for the former is higher
obscuration, then one might expect redder $V-J$ colours. The fact that
they are bluer in combination with lower optical luminosities perhaps
suggests that the AGN/SB hybrids have lower stellar masses than expected (assuming
$V$-band luminosity to be a proxy for stellar mass). Subsequently taking
the IR luminosity as a proxy for star-formation rate we infer that the AGN/SB hybrids (particularly those at low redshift)
are characterised by higher specific star-formation rates (sSFRs). This
suggests that the simultaneous black hole and
stellar mass build up in a given galaxy potentially occurs during a particularly
active phase in its star-formation history, where a large fraction of
the mass is being assembled.

\section{Summary and Conclusions}
\label{sec:conclusions}

Type-\Rmnum{2} AGN hosts and IR-luminous galaxies 
  overlap to a large extent in colour-magnitude space at green/red $U-V$ colours
  ($U-V \sim 1-1.5$) and bright $V$-band magnitudes ($M_{\rm V} \sim
  -21.5$), the latter of which is characteristic of
  massive galaxies. Large overlap between the populations is also seen in colour-colour ($U-V$
  vs $V-J$) space, with many AGN lying within the region traditionally
  occupied by dusty SFGs. We find that within
  the region of overlap, type-\Rmnum{2}
  AGN ($L_{\rm X}$\,$>$10$^{42}$\,erg/s) at $z<$1.5 are on average 3 times more
  likely (with a significance of
  $\sim$4\,$\sigma$) to reside in dusty SFGs than
  would be expected serendipitously, if black hole accretion and
  star-formation were unrelated events. This suggests that the incidence of luminous AGN in highly star-forming galaxies is not a random event; instead it indicates a link between black hole accretion and
  star-formation, likely symptomatic of favourable conditions in these
  systems, such as large gas reservoirs and the ability to funnel
  gas towards the necessary regions, e.g. via mergers, bars
  etc. Moreover we find tentative evidence that in many sources the
  AGN phase is coeval with a particularly
  active phase in the galaxy's star-formation history, during an epoch
  when a large fraction of the mass is being assembled. 

However the story does not end here, as besides significant overlap, we also note some clear differences in the
colour-magnitude and colour-colour distributions of AGN hosts and IR
galaxies. A substantial fraction of AGN hosts are offset from the parameter
space occupied by dusty galaxies, displaying bluer
$V-J$ colours closer to those expected for more quiescent,
post-starburst galaxies. These observations suggest that type-\Rmnum{2} AGN hosts are not a
  uniform population: some have colours indicative of
star-formation and dust extinction, whereas others have colours indicative of the slowing down or termination of star-formation.

Although we cannot present our results as evidence of AGN feedback, we
can nevertheless say that they are consistent with a scenario whereby AGN play a role in galaxy
evolution. Based on our observations, we propose that massive galaxies build their
black hole and stellar masses simultaneously whilst located on the
dusty galaxy sequence in colour-colour space; subsequently the AGN
potentially terminates star-formation, but outlives this event, and thus we
observe many AGN hosts in the part of colour-colour space
traditionally occupied by transitional or post-starburst systems.

\bibliographystyle{mn2e}
\bibliography{references}

\label{lastpage}

\end{document}